\documentclass[11pt,twoside]{article}

\usepackage{asp2004}
\usepackage{graphicx}

\begin{document}

\title{Detection of the Molecular Zeeman Effect in Circular Polarization 
on Cool Active Stars}

\author{S.V. Berdyugina$^{1,2}$, P. Petit$^{3}$, D.M. Fluri$^{1}$, N. Afram$^{1}$, 
J. Arnaud$^{4}$}

\affil{$^1$ Institute of Astronomy, ETH Zurich, 8092 Zurich, Switzerland\\
       $^2$ Astronomy Division, PO Box 3000, 90014 University of Oulu, Finland\\
       $^3$ Max-Planck-Institut f\"ur Sonnensystemforschung, Max-Planck-Str.~2, 
           D-37191 Katlenburg-Lindau, Germany\\
       $^4$ Laboratoire d'Astrophysique de Toulouse et Tarbes (LATT) OMP, 14, 
            Avenue Edouard Belin, F-31400 Toulouse, France 
}

%%% Abstract to run on from here.
\begin{abstract}
We report on the first ever detection of circular polarization in molecular 
lines forming in magnetic regions on the surfaces of active stars. The new 
observations were obtained with the high-resolution spectropolarimeter 
ESPaDOnS recently installed at the Canada-France-Hawaii Telescope. 
In July 2005 we carried out a survey of 17 G-K-M stars including active 
main-sequence dwarfs and RS CVn-type giants and subgiants. All stars were 
found to possess surface magnetic fields producing average atomic 
Stokes $V$ signals of 0.05\% to 0.5\%. Three stars clearly showed circular 
polarization in molecular lines of 0.5\% to 1\%. The molecular Stokes $V$ 
signal is reminiscent of that observed in sunspots. 
\end{abstract}

%%% MAIN BODY OF TEXT 

\section{Introduction}

%Studying magnetic activity on stars other than the Sun provides an 
%opportunity for detailed tests of solar dynamo models, since an extensive 
%sample of stars of various activity levels provides a wider range of
%the globals stellar parameters. A multitude of magnetic
%phenomena observed on cool active stars includes starspots in the 
%photosphere, chromospheric plages, coronal loops, UV, X-ray and radio 
%emission and flares.

Starspots are the best studied proxy of stellar magnetism. Large stellar
brightness variations and indirect imaging of stellar surfaces with the 
Doppler Imaging (DI) technique indicate immense starspot regions as compared
to sunspot sizes. Molecular lines provide an additional evidence of 
cool spots 
on the surfaces of active stars. If the effective temperature of the stellar
photosphere is high enough, molecular lines can only be formed in
cool starspots. The first detection of molecular bands from starspots was 
reported by \citet{Vogt1979} for a star whose spectral type K2 was not
compatible with the presence of TiO and VO bands. From the relative strengths 
and overall appearance of the molecular features, an equivalent spectral type 
of the spot spectrum was estimated as late as M6.

Polarimetric measurements of starspots help to investigate the nature of 
the underlying magnetic fields. However, most of our current knowledge 
about magnetic fields on cool stars and in starspots is based on Zeeman 
broadening measurements, which reveal the distribution of magnetic field 
strengths with little dependence on the unknown field geometry.
%\citep{Robinson1980, Saar1988, ValentiJohnsKrull2001}. 
An overview of the published measurements of magnetic fields on the
surfaces of cool stars \citep{Berdyugina2005} indicates a tendency for cooler
dwarfs to have stronger magnetic fields and larger areas covered by
them. Also, there is evidence that different techniques reveal different
activity signatures, such as spot umbrae and penumbrae, or even
faculae. The latter two, being brighter and possessing relatively strong
magnetic fields, would be better seen in atomic lines. This is
also supported by results obtained with the Zeeman-Doppler Imaging (ZDI) 
technique, which reveals stronger magnetic fields for intermediate 
brightness regions, and the magnetic field distribution does not
coincide with the darkest spots in the temperature images
\citep[e.g.][]{DonatiCameron1997}.
Thus, it appears that umbral magnetic fields have not been measured as
yet. 

Spectropolarimetry in molecular lines which are only formed in starspot
umbrae can provide measurements of magnetic fields directly in spatially 
unresolved spots. For example, the strongest TiO band at 7055\,\AA\ is 
magnetically quite sensitive, having effective Land{\'e} factors up to 1 
\citep{BerdyuginaSolanki2002, Berdyuginaetal2003}. Thus, a clear Stokes $V$ 
signal in the TiO band is expected from starspots \citep{Berdyugina2002, 
Aframetal2006}. Here we report the results of our spectropolarimetric 
survey of active G-K-M stars.

\section{Observations}

Observations were carried out in July, 14--16, 2005, at the Canada-France-Hawaii 
Telescope (CFHT) with the new spectropolarimeter ESPaDOnS \citep{Donati2006}. 
All measurements were 
made in the circular polarization mode with four subsequent exposures at different
polarization angles. The calibration and reduction procedures included corrections
for the dark current, flat-field, Fabry-P\'erot calibration, etc. The maximum 
polarimetric accuracy achieved was 10$^{-3}$.

Our survey included a sample of cool active stars: ten G--M dwarfs and seven G--K 
components of RS~CVn-type systems (Table~\ref{tab:obs}). The selected stars are 
moderate rotators ($v\sin i$\,$\le$\,24\,km/s) and brighter than $\sim$10th 
magnitude. All are known to have cool spots on their surfaces.

\begin{table}
\centering
\begin{tabular}{llccccr}
\hline
      Star        & Sp.\ class &  V	 &$v\sin i$& P    &$V/I_{\rm c}$&$V/I_{\rm c}$\\
                  &            &[mag]& [km/s]  &[days]& atoms       & TiO \\   
\hline
      EK Dra        & G1 V     & 7.8 &  17	& 2.6     & 0.09$^*$  &   $\le$0.2\\
      V478 Lyr      & G8 V SB1 & 7.7 &	21	& 2.15    & 0.12$^*$  &   $\le$0.1\\
      $\xi$ Boo A   & G8 V     & 4.5 &	 3	& 6.43    & 0.06      &   $\le$0.1\\
      $\xi$ Boo B   & K4 V     & 7   &	    &         & 0.06      &   $\le$0.2\\
      EQ Vir	      & K5 V     & 9.3 &  11  & 3.96    & 0.36$^*$  &   $\le$0.3\\
      BY Dra	      & K4/M0 V  & 8.1 &	8/7	& 3.83    & 0.05$^*$  &   $\le$0.2\\
      SZ UMa	      & M0e V    & 9.3 &$<$3  &         & 0.04$^*$  &   $\le$0.2\\
      AU Mic	      & M1 V     & 8.6 &	 7	& 4.85    & 0.39$^*$  &   0.4$^*$\\
      FK Aqr	      & M2/M3e V & 9.1 &	  	& 4.39    & 0.30$^*$  &   0.5$^*$\\
      EV Lac	      & M3.5e V  &10.1 &   7  & 4.38    & 0.28$^*$  &   1.1$^*$\\
\hline
      $\lambda$ And & G8 IV	   & 3.9 &	10	& 53.95   & 0.09$^*$  &   $\le$0.1\\
      HK Lac	      & K0 III	 & 6.8 &  24  & 24.46   & 0.08$^*$  &   $\le$0.1\\
      XX Tri	      & K0 III	 & 8.7 &	17	& 24      & 0.14$^*$  &   $\le$0.2\\
      29 Dra	      & K1 III	 & 6.7 &	 8	& 31.5    & 0.14$^*$  &   $\le$0.1\\
      BM CVn	      & K1 III	 & 7.4 &	15	& 20.6    & 0.19$^*$  &   $\le$0.1\\
      IM Peg	      & K1 III	 & 5.8 &	24	& 24.65   & 0.05      &   $\le$0.1\\
      II Peg	      & K2 IV	   & 7.7 &	22	& 6.7     & 0.36      &   $\le$0.2\\
\hline
\end{tabular}
\caption{Observed targets. The last two columns provide peak circular 
polarization (\%) in atomic and molecular lines measured from LSD profiles 
and the TiO 7055\,\AA\ band, respectively. Upper limits are estimated from
the noise level. Asterisks mark cases with first detections.}
\label{tab:obs}
\end{table}

\section{Results}

\begin{figure}
\begin{centering}
\includegraphics[width=13cm]{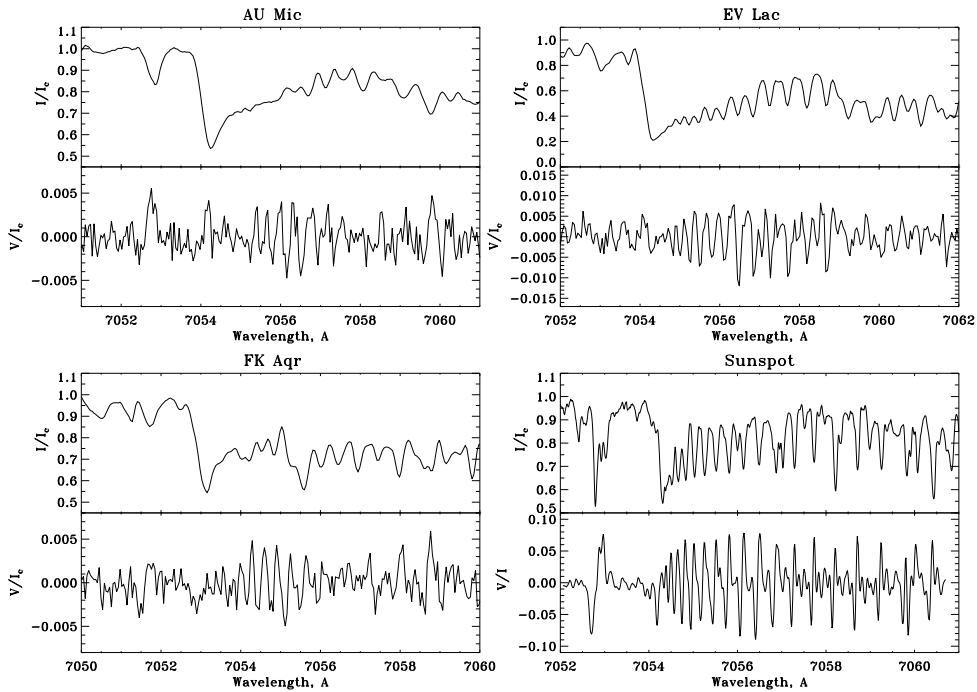}
\caption{Observed Stokes $I/I_{\rm c}$ (upper panels) and Stokes $V/I_{\rm c}$ 
(lower panels) of the TiO $\gamma$\,(0,0) $R_3$ band head on the active M dwarfs 
AU~Mic, FK~Aqr and EV~Lac. Note a close similarity to the Stokes signals in 
a sunspot.
\label{fig:tio}}
\end{centering}
\end{figure}

A clear Stokes $V$ signal in the TiO 7055\AA\ band (up to 1\%) was 
detected on three M dwarfs (Fig.~\ref{fig:tio}). Two stars (AU~Mic and EV~Lac) 
were known to have strong ($\sim$4\,kG) surface magnetic fields measured from 
Zeeman-broadened atomic lines \citep{Saar1992, JohnsKrullValenti1996}. 
It is however the first detection of magnetic fields on both components of FK~Aqr.
A simple modeling of the observed circular polarization indicates single-polarity 
magnetic fields on the three stars covering at least 10\% of the stellar
disk \citep{Aframetal2006}.

No Stokes $V$ signal in the TiO band above the noise level of 0.1--0.2\% was 
detected on the G--K dwarfs and giants from our list (Table~\ref{tab:obs}). 
This implies that the magnetic fields on these stars are apparently weaker,
more entangled than on M dwarfs, or more diluted by the bright photosphere.

\begin{figure}
\begin{centering}
\includegraphics[width=13cm]{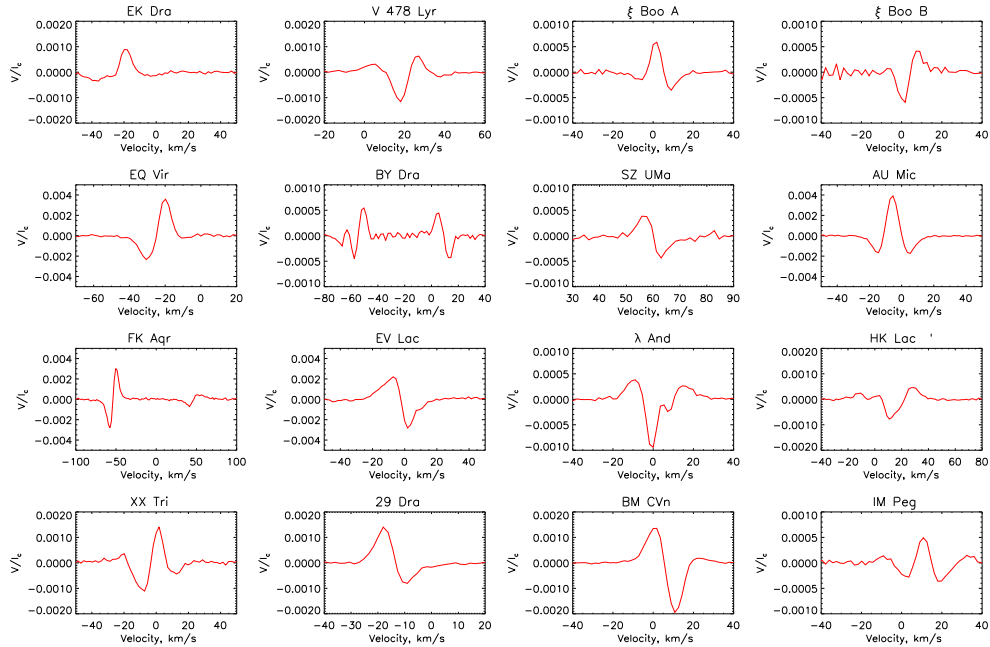}
\caption{Observed atomic LSD Stokes $V/I_{\rm c}$ profiles.
\label{fig:lsd}}
\end{centering}
\end{figure}

A magnetic field was actually detected on all stars in average atomic Stokes 
$V$ profiles extracted with the Least Squares Deconvolution (LSD) technique 
\citep{Donatietal1997}. For most stars this is the first detection 
(Table~\ref{tab:obs}, Fig.~\ref{fig:lsd}). Note that the largest signals in the atomic
Stokes $V$ profiles were observed in the three coolest M dwarfs, where the TiO Stokes $V$
signals were prominent as well, and in the two coolest K stars EQ~Vir and II~Peg, where
perhaps the higher noise level prevented detection of TiO. On all these stars Stokes $V$
profiles in individual atomic lines were also recorded. A simultaneous analysis of 
the Stokes $I$ and $V$ signals from many atomic and molecular lines with 
different temperature and magnetic sensitivities will allow us to disentangle the 
contributions from the photosphere, faculae and starspot umbrae and penumbrae.

%%% Text of acknowledgements runs on after this command.
\acknowledgements 
Our observations were successful thanks to the excellent team work together with the CFHT staff, our support astronomer Nadine Manset, and of course ESPaDOnS and its creators (PI J.-F. Donati). We are also thankful to the OPTICON access programme for the support of our observations.
This work was supported by ETH Research Grant TH-2/04-3 and SNF grants PE002-104552 and 200021-103696. S.V.~Berdyugina acknowledges 
the EURYI award from the ESF.

%%% THE BIBLIOGRAPHY

\end{document}